\documentclass[3p,times,twocolumn]{elsarticle}

\usepackage{ecrc}
\usepackage{amsmath}
\usepackage[utf8]{inputenc}
\usepackage{slashed}

\volume{00}

\firstpage{1}

\journalname{Nuclear Physics B Proceedings Supplement}

\runauth{}

\jid{nuphbp}

\jnltitlelogo{Nuclear Physics B Proceedings Supplement}

\usepackage{amssymb}

\usepackage[figuresright]{rotating}

\begin{document}

\begin{frontmatter}



\dochead{}

\title{Kaon semileptonic vector form factor with $N_f=2+1+1$ Twisted Mass fermions}


\author[a]{N. Carrasco}
\author[a,b,c]{P. Lami}
\author[a,b]{V. Lubicz}
\author[a,b]{E. Picca}
\author[a]{L. Riggio}
\author[a]{S. Simula}
\author[a,b]{C. Tarantino}

\address[a]{INFN sezione Roma Tre, Via della Vasca Navale 84 00154, Roma, Italia}
\address[b]{Dipartimento di Matematica e Fisica Università degli Studi Roma Tre, Via della Vasca Navale 84 00154, Roma, Italia}
\address[c]{Presenter}
\begin{abstract}
We investigate the vector form factor relevant for the $K_{\ell 3}$ semileptonic decay using maximally twisted-mass fermions with 4 dynamical flavours ($N_f=2+1+1$).
Our simulations feature pion masses ranging from $210$ MeV to approximately $450$ MeV and lattice spacing values as small as $0.06$fm. Our main result for the vector form factor at zero 4-momentum transfer is $f_+(0)=0.9683(65)$ where the uncertainty is both statistical and systematic.
By combining our result with the experimental value of $f_+(0)|V_{us}|$ we obtain $|V_{us}|=0.2234(16)$, which satisfies the unitarity constraint of the Standard Model at the permille level.
\end{abstract}

%
%

\end{frontmatter}


\section{Introduction and simulation details}
\label{sec:int}
Meson semileptonic decays are very interesting phenomena because the measure of their rates combined with lattice QCD calculations allows us to extract the elements of the CKM matrix. Specifically, in this contribution we present our calculation of the vector form factor of the kaon semileptonic decay at zero 4-momentum transfer, which allows us to extract
the value of $|V_{us}|$. 

This is possible thanks to the relation between the vector current responsible for the decay and two form factors:
\begin{equation}
 \left<  \pi(p')|V_{\mu}|K(p)  \right>=(p_\mu+p'_\mu)f_+(q^2)+(p_\mu-p'_\mu)f_{-}(q^2),
\end{equation}
where $q_\mu=p_\mu-p'_\mu$.
The definition of the scalar form factor $f_0$ is
\begin{equation}
\label{eq:f0def}
f_0(q^2)=f_+(q^2)+\frac{q^2}{M_K^2-M_{\pi}^2}f_{-}(q^2),
\end{equation}
which implies the relation $f_+(0)=f_0(0)$.
By calculating $f_+(0)$ on the lattice and using the experimental result of $f_+(0)|V_{us}|$ we can then extract the value of $|V_{us}|$.

We used the ensembles produced by the ETM Collaboration with $N_f=2+1+1$ using the Twisted Mass action \cite{Frezzotti:2003zc,Frezzotti:2003ni}, which include in the sea, beside the contribution of two degenerate light quarks, the strange and charm quarks. Our simulation features pion masses ranging from $210$ MeV to approximately $450$ MeV and three values of the lattice spacing, the smallest being approximately $0.06$fm. Valence quarks were simulated using the Osterwalder-Seiler action \cite{Osterwalder:1977pc}, while the gauge fields were implemented using the Iwasaki action \cite{Iwasaki:1985we}.
For each lattice spacing we used three values of the bare strange quark mass to allow for a smooth interpolation of our data to the physical value $m_s$, which we determined in our paper \cite{Carrasco:2014cwa}.
Different values of the spatial momenta were simulated using Twisted Boundary conditions \cite{Bedaque:2004kc,deDivitiis:2004kq}, allowing us to cover both the spacelike and timelike region of the 4-momentum transfer.
For further details about the simulation the reader should see ref. \cite{Carrasco:2014cwa}.

We studied a combination of three-points correlation functions in order to extract the form factors $f_+$ and $f_0$ as functions of the 4-momentum transfer $q^2$, light quark mass $m_\ell$ and the lattice spacing $a$.
We then performed a chiral and continuum extrapolation in order to obtain the physical value of $f_+(0)$.

Our result is $f_+(0)=0.9683(65)$ where the uncertainty is both statistical and systematic.This allows us to extract the value of the CKM matrix element $|V_{us}|=0.2234(16)$, which is compatible with the unitarity constraint of the Standard Model at the permille level.

\section{Extraction of the form factors}
Our data consists of three point correlation functions connecting moving pions and kaons through a vector current inserted at a time distance $t$ from the source and $(T/2-t)$ from the sink.
The behaviour of these correlation functions for large $t/a$ allows us to extract the matrix elements of the vector current by studying the following quantity
\begin{equation}
\begin{split}
R_\mu(t,\vec{p},\vec{p'})&=\frac{ C_{\mu}^{K\pi}(t,\vec{p},\vec{p'})C_{\mu}^{\pi K}(t,\vec{p'},\vec{p})}{C_{\mu}^{\pi\pi}(t,\vec{p'},\vec{p'})C_{\mu}^{KK}(t,\vec{p},\vec{p})},  \\
R_\mu\xrightarrow{\scriptscriptstyle\substack{t\to\infty}}&\frac{\left<  \pi(p')|V_{\mu}|K(p)  \right> \left< K(p) |V_{\mu}| \pi(p') \right>}{\left<  \pi(p')|V_{\mu}| \pi(p') \right>\left< K(p) |V_{\mu}|K(p) \right>}.
\end{split}
\end{equation}
The matrix elements $\left<V_0\right>$ and $\left<V_i\right>$ can be extracted from the $R_\mu(t,\vec{p},\vec{p'})$ plateaux as
\begin{equation}
\begin{split}
&\left<  \pi(p')|V_{0}|K(p)  \right>=\left< V_0 \right>=2\sqrt{R_0}\sqrt{EE'},\\
&\left<  \pi(p')|V_{i}|K(p)  \right>=\left< V_i \right>=2\sqrt{R_i}\sqrt{p_ip_i'}.
\end{split}
\end{equation}
Thus, we obtain the form factors through the relations
\begin{equation}
\begin{split}
&f_{+}(q^{2})=\frac{(E-E')\left<V_i\right>-(p_i-p'_i)\left<V_0\right>}{2Ep'_i-2E'p_i},\\
&f_{-}(q^{2})=\frac{(p_i+p'_i)\left<V_0\right>-(E+E')\left<V_i\right>}{2Ep'_i-2E'p_i},
\end{split}
\end{equation}
and subsequentely calculate $f_0(q^2)$ from eq.(\ref{eq:f0def}).
\begin{figure}
\scalebox{0.3}{\includegraphics{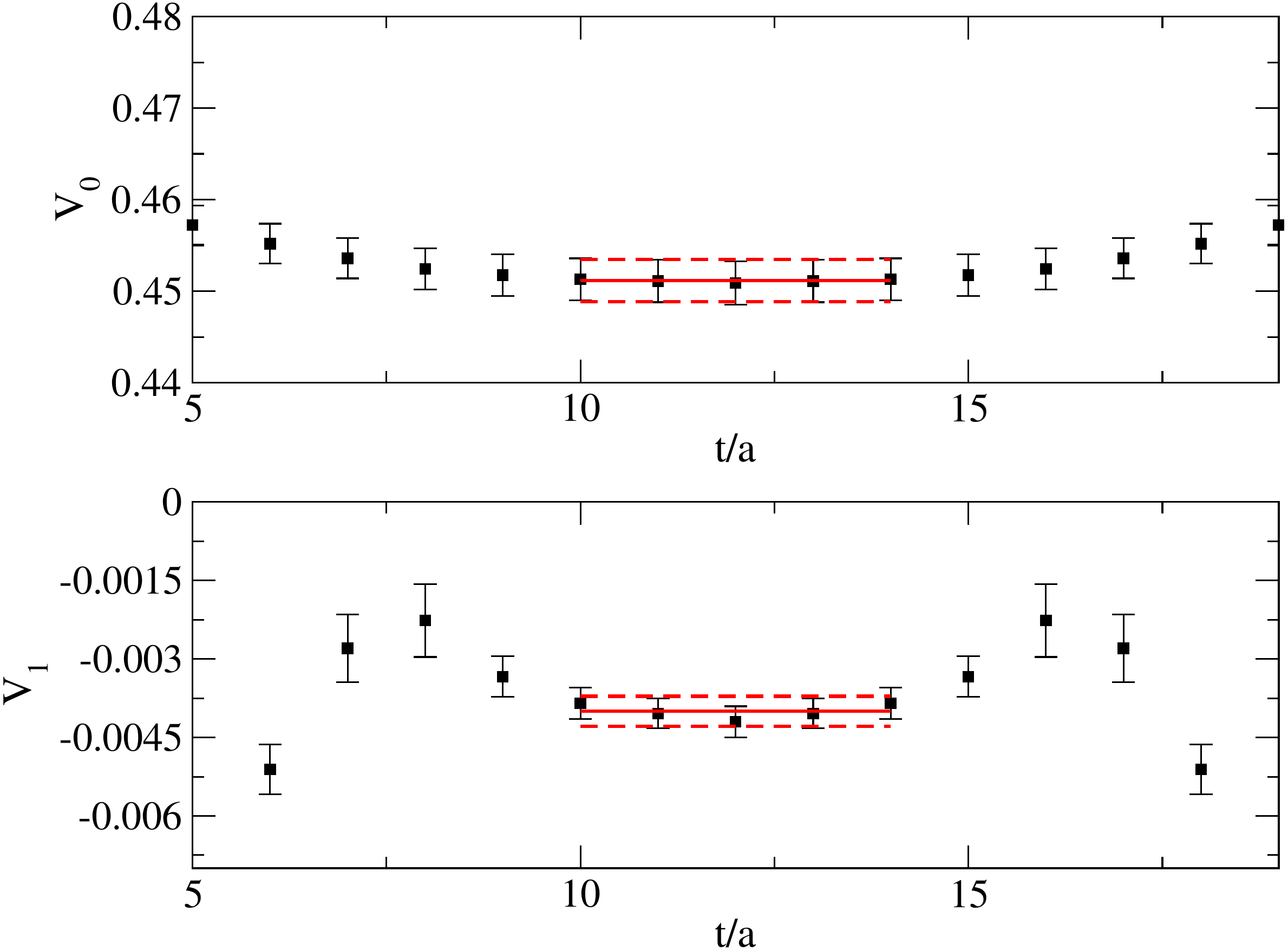}}
\caption{Example of the matrix elements $\left< V_0 \right>$ and $\left< V_i \right>$ extracted from the quantity $R_{\mu}$ corresponding to an ensemble with $\beta=1.90$, $L/a=24$, $a\mu_l=0.0080$, $a\mu_s=0.0225$, $|\vec{p}|=|\vec{p'}|\simeq87$MeV. }
\label{fig:matel}
\end{figure}
An example of the extraction of the matrix elements can be seen in fig.(\ref{fig:matel}).
Meson masses were calculated by isolating the ground state of two points correlation functions of pseudoscalar mesons at rest. 
\section{Analysis of the form factors}
\begin{figure}
\scalebox{0.3}{\includegraphics{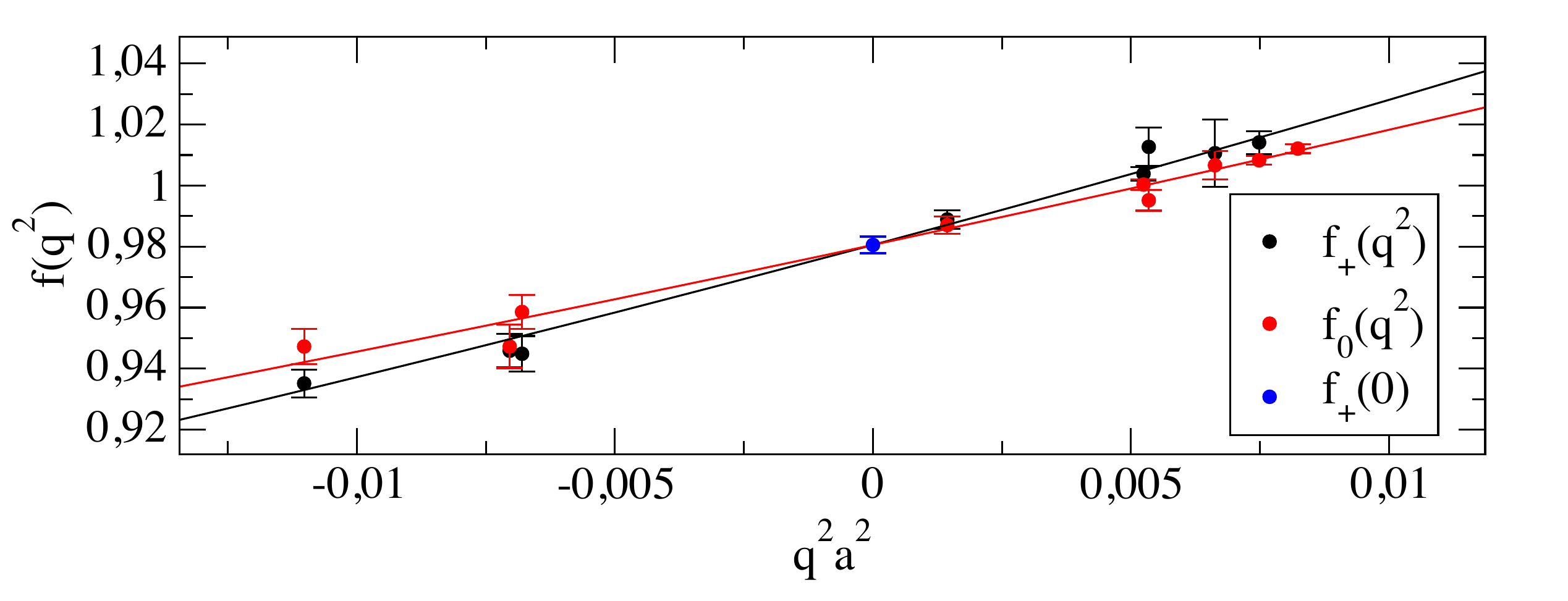}} 
\vspace{-0.5cm}
\caption{Interpolation of the form factors $f_+(q^2)$ and $f_0(q^2)$ data to $q^2=0$ using the $z$ expansion \cite{Hill:2006bq}. The plot refers to an ensemble with $\beta=1.90$, $L/a=24$, $a\mu_l=0.0060$, $a\mu_s=0.0225$. }
\label{fig:f0fpq2}
\end{figure}

The first step in our analysis was to study the dependence of $f_+$ and $f_0$ on the 4-momentum transfer $q^2$ in order to interpolate our data to $q^2=0$. This was done using the $z$ expansion \cite{Hill:2006bq} (up to $\mathcal{O}(z)$) and the condition $f_+(0)=f_0(0)$ is imposed as a constraint. An example can be seen in fig.(\ref{fig:f0fpq2}). We also tried to fit the $q^2$ dependence using other fit ansatz (e.g. polynomial expression in $q^2$), obtaining nearly identical results.
Thus, after interpolating our data to the physical value of the strange quark mass using a quadratic spline procedure, we performed the chiral and continuum extrapolation using the following SU(2) ChPT prediction at NLO \cite{Flynn:2008tg}:
\begin{equation}
\label{eq:su2}
 f_{+}(0)=F^+_0\left( 1-\frac{3}{4}\xi_l \log\xi_l+P_2\xi+P_3a^2 \right),
\end{equation}
where $xi=2B_0m_\ell/(4\pi f_0)^2$, $a$ is the lattice spacing and the parameters $F^+_0$, $P_2$ and $P_3$ are determined by our fit.

\begin{figure}
\scalebox{0.3}{\includegraphics{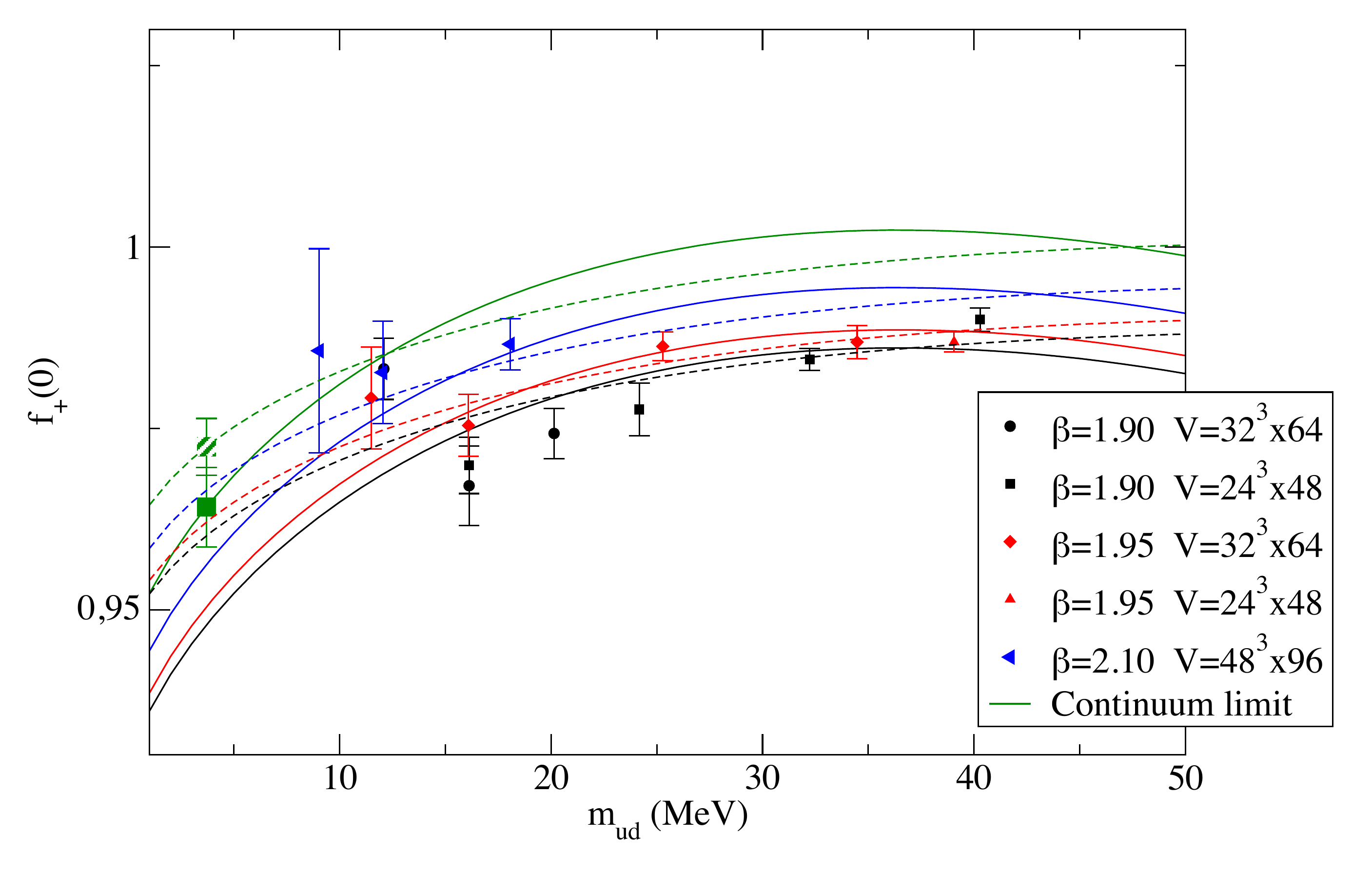}}
\caption{Chiral and continuum extrapolation of the quantity $f_+(0)$, showing both the SU(2) (solid lines) and the SU(3) (dashed lines) ChPT fit results}
\label{fig:f0ml}
\end{figure}

In order to estimate the systematic uncertainty induced by the chiral extrapolation we also fitted our data using the SU(3)  ChPT ansatz beyond the NLO:
\begin{equation}
\label{eq:su3}
 f_{+}(0)=1+f_2+(M_K^2-M_\pi^2)^2[\Delta_1+(M_K^2+M_\pi^2)\Delta_2]+\Delta_3a^2,
\end{equation}

where $\Delta_1$, $\Delta_2$ and $\Delta_3$ are determined in our fit. The full expression for $f_2$ can be found in \cite{Gasser:1984gg,Gasser:1984ux}.
In eq.(\ref{eq:su3}) the Ademollo Gatto theorem \cite{Ademollo:1964sr} is satisfied in the continuum limit, i.e. in the SU(3) limit  $f_+(0)=1$ and the deviations from this value are quadratic in $(M_K^2-M_\pi^2)$.
In fig.(\ref{fig:f0ml}) we show the chiral and continuum extrapolation of our data, using eqs. (\ref{eq:su2}) and (\ref{eq:su3}). It can be seen that the results at the physical point are compatible within the uncertainties.
Thus we combined the two results obtaining 
\begin{equation}
\label{eq:fp0res}
f_+(0)=0.9683(50)_{stat+fit}(42)_{Chir}=0.9683(65),
\end{equation}
where $()_{stat+fit}$ indicates the statistical uncertainty which includes the one induced by the fitting procedure and the error induced by the numerical inputs needed for the analysis, namely the values of the light quark mass $m_\ell$, the lattice spacing $a$ and the SU(2) ChPT low energy constants $f_0$ and $B_0$, which were determined in \cite{Carrasco:2014cwa}.
The $()_{Chir}$ part of the uncertainty is the one induced by the difference in the results corresponding to the two chiral extrapolations we performed.
It should be noticed that there are two lattice points calculated at the same lattice spacing and light quark mass but different volumes. They turn out to be well compatible within the uncertainties, allowing us to state that finite size effects can be safely neglected in our analysis.

We then took the experimental value of $|V_{us}|f_+(0)$ from \cite{Antonelli:2010yf} and obtained 
\begin{equation}
|V_{us}|=0.2234(16). 
\end{equation}
Taking the result for $|V_{ud}|$ from \cite{Hardy:2009} we perform the unitarity test
\begin{equation}
\label{eq:utest}
 |V_{ud}|^2+|V_{us}|^2+|V_{ub}|^2=0.9991(8),
\end{equation}
where the contribution of $|V_{ub}|^2$ is negligible.

\section{An outlook on a possible extension }
As a possible extension of our analysis we performed a multi-combined fit of the $q^2$, $m_\ell$ and $a$ dependencies of the form factors in order to predict them not only at $q^2=0$, but on the entire $q^2$ region accessible to experiments, i.e from $q^2=0$ to $q^2=q^2_{max}=(M_K-M_\pi)^2$

\begin{figure}
\scalebox{0.3}{\includegraphics{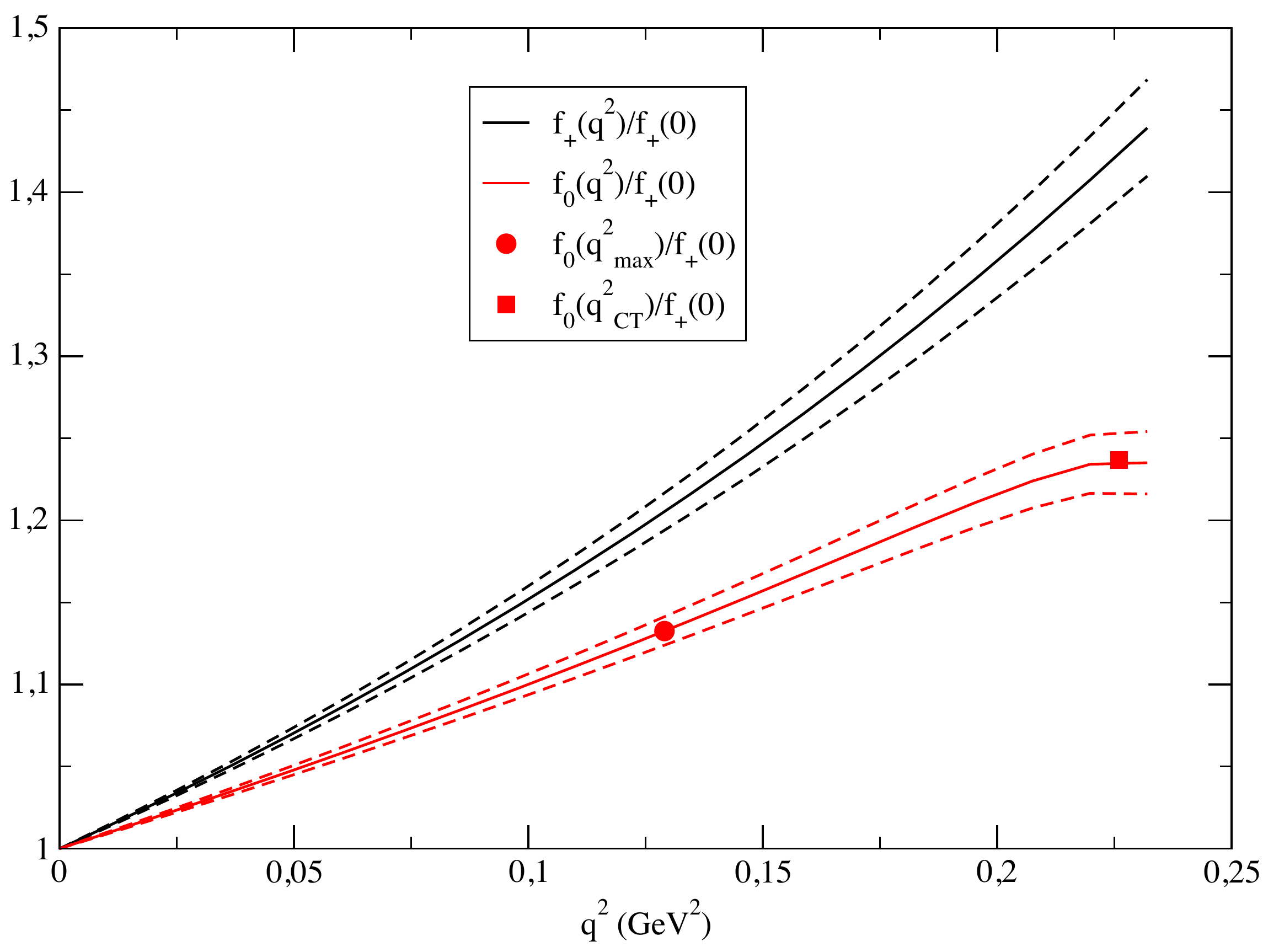}}
\caption{Fit results for the quantities $f_+(q^2)/f_+(0)$ and $f_0(q^2)/f_+(0)$ as functions of $q^2$ at the physical point. The red dot (square) corresponds to $q^2_{max}$  ($q^2_{CT}$).}
\label{fig:f0fpoverfp0}
\end{figure}

We opted for the same strategy used in \cite{Lubicz:2010bv}, i.e. we performed a global fit using functional forms of the form factors derived by expanding in powers of $x=M_{\pi^2}/M_K^2$ the NLO SU(3) ChPT predictions for the form factors \cite{Gasser:1984ux,Gasser:1984gg}. As a constraint we included in the analysis the Callan-Treiman theorem \cite{Callan:1966hu}, which relates in the SU(2) chiral limit the scalar form factor calculated at the unphysical $q^2_{CT}=M_K^2-M_\pi^2$ to the ratio of the decay constants $f_K/f_\pi$.
A preliminary result for the form factors is presented in fig.(\ref{fig:f0fpoverfp0}).
\vspace{-0.4cm}
\section{Aknowledgements}
The authors would like to thank Francesco Sanfilippo for the useful discussions about the work here presented.




\nocite{*}
\bibliographystyle{elsarticle-num}

\vspace{-0.3cm}
\begin{thebibliography}{00}
 
 
%
\bibitem{Frezzotti:2003zc}
  R.~Frezzotti and G.~C.~Rossi,
  Nucl.\ Phys.\ Proc.\ Suppl.\  {\bf 129} (2004) 880
  [hep-lat/0309157].
 
%
\bibitem{Frezzotti:2003ni}
  R.~Frezzotti and G.~C.~Rossi,
  JHEP {\bf 0408} (2004) 007
  [hep-lat/0306014].
 
%
\bibitem{Osterwalder:1977pc}
  K.~Osterwalder and E.~Seiler,
  Annals Phys.\  {\bf 110} (1978) 440.
 
 
%
\bibitem{Iwasaki:1985we}
  Y.~Iwasaki,
  Nucl.\ Phys.\ B {\bf 258} (1985) 141.
  
 
 \bibitem{Carrasco:2014cwa}
  N.~Carrasco {\it et al.},
  Nucl.\ Phys.\ B {\bf 887} (2014) 19
  [arXiv:1403.4504 [hep-lat]].
%
\bibitem{Bedaque:2004kc}
  P.~F.~Bedaque,
  Phys.\ Lett.\ B {\bf 593} (2004) 82
  [nucl-th/0402051].

  
%
\bibitem{deDivitiis:2004kq}
  G.~M.~de Divitiis, R.~Petronzio and N.~Tantalo,
  Phys.\ Lett.\ B {\bf 595} (2004) 408
  [hep-lat/0405002].
  
  

  
 

\bibitem{Hill:2006bq}
  R.~J.~Hill,
  Phys.\ Rev.\ D {\bf 74} (2006) 096006
  [hep-ph/0607108].
 
%
\bibitem{Flynn:2008tg}
  J.~M.~Flynn {\it et al.}  [RBC and UKQCD Collaborations],
  Nucl.\ Phys.\ B {\bf 812} (2009) 64
  [arXiv:0809.1229 [hep-ph]].
 
 
\bibitem{Gasser:1984ux}
  J.~Gasser and H.~Leutwyler,
  Nucl.\ Phys.\ B {\bf 250} (1985) 517.
 
\bibitem{Gasser:1984gg}
  J.~Gasser and H.~Leutwyler,
  Nucl.\ Phys.\ B {\bf 250} (1985) 465.
%
\bibitem{Ademollo:1964sr}
  M.~Ademollo and R.~Gatto,
  Phys.\ Rev.\ Lett.\  {\bf 13} (1964) 264.
  
%
\bibitem{Antonelli:2010yf}
  M.~Antonelli {\it et al.},
  Eur.\ Phys.\ J.\ C {\bf 69} (2010) 399
  [arXiv:1005.2323 [hep-ph]].
  
\bibitem{Hardy:2009}
  J. C. Hardy and I. S. Towner,
  Phys.\ Rev.\ C {\bf 79} (2009) 055502
  [hep-ph/0607108].
  
%
\bibitem{Lubicz:2010bv}
  V.~Lubicz {\it et al.}  [ETM Collaboration],
  PoS LATTICE {\bf 2010} (2010) 316
  [arXiv:1012.3573 [hep-lat]].
  
  

  
%


  
%
\bibitem{Callan:1966hu}
  C.~G.~Callan and S.~B.~Treiman,
  Phys.\ Rev.\ Lett.\  {\bf 16} (1966) 153.
  
\end{thebibliography}







\end{document}